# An Efficient PTAS for Two-Strategy Anonymous Games

Constantinos Daskalakis
Microsoft Research*

**Abstract**

We present a novel polynomial time approximation scheme for two-strategy anonymous games, in which the players' utility functions, although potentially different, do not differentiate among the identities of the other players. Our algorithm computes an $\epsilon$-approximate Nash equilibrium of an $n$-player 2-strategy anonymous game in time $\text{poly}(n) \cdot (1/\epsilon)^{O(1/\epsilon^2)}$, which significantly improves upon the running time $n^{O(1/\epsilon^2)}$ required by the algorithm of Daskalakis & Papadimitriou, 2007. The improved running time is based on a new structural understanding of approximate Nash equilibria: We show that, for any $\epsilon$, there exists an $\epsilon$-approximate Nash equilibrium in which either only $O(1/\epsilon^3)$ players randomize, or all players who randomize use the same mixed strategy. To show this result we employ tools from the literature on Stein's Method.

## 1 Introduction

It has been recently established that computing a Nash equilibrium is an intractable problem [18, 10, 6, 13], even in the case of two-player games [7]. In view of this hardness result, research has been directed towards the computation of approximate Nash equilibria, which are states of the game in which no player has more than some small $\epsilon$ incentive to change her strategy. But, despite much research in this direction [22, 21, 11, 17, 12, 5, 27], only constant $\epsilon$'s can be achieved in polynomial time. Yet, an approximate Nash equilibrium in which the players have regret equal to a significant fraction of their payoffs is not an attractive solution concept; after all, there is no reason to expect a player to keep her strategy if she can significantly improve by changing to a different one. On the contrary, if $\epsilon$ were arbitrarily small, it could be that the cost of switching one's strategy is larger than the regret $\epsilon$ that she suffers. Hence, approximate equilibria with arbitrarily close approximation could be credible solutions concepts. The following question then emerges: *Is there a Polynomial Time Approximation Scheme for approximate Nash equilibria?*

The question remains open for general games, but there are special classes known to be tractable. It is well-known, for example, that zero-sum games are solvable exactly in polynomial time by Linear Programming [24, 9]. This tractability result has been extended to a generalization of zero-sum games, called *two-player low-rank games*, in which the sum of the players' payoff tables has fixed rank; in this case there is a PTAS for approximate Nash equilibria. It has also been shown that symmetric multi-player games with (about logarithmically) few strategies per player can be solved exactly in polynomial time by a reduction to the theory of real closed fields [25]. In congestion games, we can compute in polynomial time a pure Nash equilibrium, if the game is a symmetric network congestion game [16], and an approximate pure Nash equilibrium, if the congestion game is symmetric (but not necessarily network) and the utilities are somehow "continuous" [8].

---

*This research was done while the author was a student at UC Berkeley. Supported by a Microsoft Research Fellowship, NSF grant CCF - 0635319 and a MICRO grant.

In this paper, we consider another important class of games, called *anonymous*. These are games in which each player's utility function does not differentiate among the identities of the other players. That is, the payoff of a player depends on the strategy that she chooses and only the *number* of other players choosing each strategy. Anonymous games comprise a broad and well studied class of games (see, e.g., [3, 4, 19, 23] for recent work on this subject by economists) which are of special interest to the Algorithmic Game Theory community, as they capture important aspects of auctions and markets, as well as of Internet congestion.

But, what do we know about computing Nash equilibria in anonymous games? It was recently established that there is a PTAS for the case of a constant number of strategies per player [14, 15]. The running time of the algorithm given in [15] is $n^{O(f(s,1/\epsilon))}$, where $\epsilon$ is the desired approximation, $s$ the number of strategies available to the players, and $f$ some function which is polynomial in $1/\epsilon$, but superpolynomial in $s$. Hence, although theoretically efficient for any fixed $\epsilon$ and $s$, the algorithm is highly non-practical. Even for the simpler case of two-strategy anonymous games the running time achieved by [14] is $n^{O(1/\epsilon^2)}$.

In this paper, we present a more efficient algorithm for 2-strategy anonymous games, which runs in time $\text{poly}(n) \cdot (1/\epsilon)^{O(1/\epsilon^2)}$. The improved running time is due to a novel understanding of certain structural properties of approximate Nash equilibria. In particular, we show that, for any integer $k$, there exists an $\epsilon$-approximate Nash equilibrium, with $\epsilon = O(1/k)$, in which

(a) either at most $k^3 = O((1/\epsilon)^3)$ players use randomized strategies, and their strategies are integer multiples of $1/k^2$; [1]

(b) or all players who randomize choose the same mixed strategy which is also an integer multiple of $\frac{1}{kn}$.

To derive the above characterization, we study mixed strategy profiles in the proximity of a Nash equilibrium. We establish that there always exists a nearby mixed strategy profile which is of one of the types (a) or (b) described above and satisfies the Nash equilibrium conditions to within an additive $\epsilon$, thus corresponding to an $\epsilon$-approximate equilibrium. Given this structural result (see Theorem 3.1), an $\epsilon$-approximate equilibrium can be found by dynamic programming (see Theorem 3.2).

We feel that a more sophisticated analysis can establish similar structural properties for approximate Nash equilibria in multi-strategy anonymous games, extending our efficient PTAS to anonymous games with any fixed number of strategies.

**Overview of Techniques:** Let us track down the effect in the Nash equilibrium resulting by replacing a mixed Nash equilibrium $(p_1, \ldots, p_n) \in [0, 1]^n$ by another strategy profile $(q_1, \ldots, q_n) \in [0, 1]^n$, where the probabilities $p_i$ and $q_i$ correspond to the mixed strategy of player $i$ in the two strategy profiles. It is not hard to see that the approximation achieved by the strategy profile $(q_1, \ldots, q_n)$ can be, loosely speaking, bounded by the total variation distance between the distribution of the sum of $n$ Bernoulli random variables with expectations $p_1, \ldots, p_n$ and that of another sum of Bernoulli random variables with expectations $q_1, \ldots, q_n$. Hence, to establish the structural property described above it is sufficient to show that given any set of probability values $(p_1, \ldots, p_n)$ there is another set $(q_1, \ldots, q_n)$ which satisfies either Property (a) or Property (b) and is such that the total variation distance between the two sums of Bernoulli random variables with expectations $\{p_i\}_i$ and $\{q_i\}_i$ respectively is at most $\epsilon$.

---

[1] Note that, since every player has 2 strategies, a mixed strategy is a number in $[0, 1]$.

To give some insight into the construction of the set $\{q_i\}_i$, let us consider the following scenarios for an integer $k$:

(i) at least $k^3$ of the $p_i$'s fall in the set $[1/k, 1 - 1/k]$ and the others are either 0 or 1;

(ii) at most $k^3$ of the $p_i$'s fall in the set $[1/k, 1 - 1/k]$ and the others are either 0 or 1;

Let us consider Case (i) first. It is reasonable to expect, by the Central Limit Theorem, or finitary versions thereof, that the sum of at least $k^3$ Bernoulli random variables with expectations from the set $[1/k, 1 - 1/k]$ is close in total variation distance to a Normal distribution, with the appropriate mean and variance, and, hence, to a Binomial distribution which approximates that Normal distribution. In Section 6.1 we show that this is indeed the case, even if we only allow the probability of the Binomial distribution to be an integer multiple of $1/kn$. Hence, the sum of the original Bernoulli random variables with expectations from the set $[1/k, 1 - 1/k]$ can be approximated by another set of Bernoulli random variables which all share the same mean, which, moreover, is an integer multiple of $1/kn$. Hence, an approximate equilibrium satisfying Property (b) above can be defined.

In the Case (ii), approximating by a Normal distribution is not tight enough to give overall total variation distance of $O(1/k)$. We resort instead to the following structural result shown in [14]: Given any set of Bernoulli random variables with expectations $p_1, \ldots, p_n$, there is a way to round the probabilities to multiples of $1/k^2$, for any $k$, so that the distribution of the sum of these $n$ variables is affected by an additive $O(1/k)$ in total variational distance. Hence, an approximate equilibrium satisfying Property (a) above can be defined (see Section 6.2).

It remains to argue that the Cases (i) and (ii) are general enough. For this, we describe an iterative procedure which alters the values of those $p_i$'s falling outside the set $\{0,1\} \cup [1/k, 1-1/k]$ in such a way that, in the end of this procedure, all $p_i$'s *do* fall in the set $\{0, 1\} \cup [1/k, 1 - 1/k]$ and the distribution of the sum of Bernoulli random variables with expectations $p_i$ does not change by more than $O(1/k)$ in total variation distance. Let us, e.g., consider the $p_i$'s falling in the set $(0, 1/k)$ and round some of them to 0 and some of them to $1/k$ so that their sum is approximated to within $O(1/k)$. By the law of rare events (the precise result we use is from the literature on Poisson approximations, see, e.g., [1]), the sum of Bernoulli's before and after the rounding is distributed like two Poisson distributions with means equal to the sum of $p_i$'s before and after the rounding respectively. Since these means are within $O(1/k)$, the two Poisson distributions are within $O(1/k)$ in total variation distance and so are the sums of the Bernoulli's before and after the rounding. For details see Section 5.

## 2 Definitions and Notation

A game has $n$ players, $1, \ldots, n$, and $t$ strategies, $1, 2, \ldots, t$, available to them, so that each player gets some payoff for every selection of strategies by her and the other players. The game is called *anonymous*, if the payoff of each player depends on her strategy and only the *number*, but not the identities, of the other players who choose each of the $t$ strategies.

In this paper, we study two-strategy anonymous games. In these games, the payoff function of each player $i$ is specified by giving $u_1^i, u_2^i : \{0, 1, \ldots, n-1\} \to [0,1]$,[2] so that $u_s^i(m)$ is the payoff of $i$, if she chooses strategy $s$ and $m$ of the other players choose strategy 2. Hence, the game is *succinctly representable* [25], in the sense that its representation requires $2n^2$ numbers, as opposed to the (exponential in the number of players) $nt^n$ numbers required for general games. Arguably, succinct games are the only multiplayer games that are computationally meaningful (see [25] for an extensive discussion of this point).

---
[2] In the literature on Nash approximation, utilities are usually normalized in this way so that the approximation error is additive.

A *mixed strategy profile* is a set of $n$ probability values $p_1, p_2, \ldots, p_n \in [0, 1]$, corresponding to the probability with which each player chooses strategy 2. A mixed strategy profile is an $\epsilon$-*Nash equilibrium* if, for all $i \in [n]$, the following hold

$$E_{\{p_j\}_{j \neq i}} u_1^i(x) > E_{\{p_j\}_{j \neq i}} u_2^i(x) + \epsilon \Rightarrow p_i = 0,$$
$$E_{\{p_j\}_{j \neq i}} u_2^i(x) > E_{\{p_j\}_{j \neq i}} u_1^i(x) + \epsilon \Rightarrow p_i = 1,$$

where for the purposes of the expectation $x$ is drawn from $\{0, \ldots, n-1\}$ by tossing $n-1$ independent coins with probabilities $\{p_j\}_{j \neq i}$. That is, a mixed strategy profile is an $\epsilon$-Nash equilibrium if every player is only randomizing among strategies which, when played against the mixed strategies of the other players, achieve expected payoff within (additive) $\epsilon$ from the expected payoff achieved by the best strategy.

The notion of $\epsilon$-Nash equilibrium is closely related to the notion of $\epsilon$-*approximate Nash equilibrium*, defined as any mixed strategy profile in which no player can improve his expected payoff by more than $\epsilon$ by changing to a different mixed strategy. It is easy to see that any $\epsilon$-Nash equilibrium is also an $\epsilon$-approximate Nash equilibrium, but the opposite implication is not true in general. In this paper, we present algorithms for computing (the stronger notion of) $\epsilon$-Nash equilibria.

Anonymous games can be extended to ones in which there is also a finite number of *types* of players, and utilities depend on how many players of each type play each of the available strategies. Our algorithm can be easily generalized to this framework, with the number of types multiplying the exponent of the running time.

Let us conclude this section with a few more definitions. We define the *total variation distance* between two distributions $\mathbb{P}$ and $\mathbb{Q}$ supported on a finite set $\mathcal{A}$ as follows

$$||\mathbb{P} \; ; \; \mathbb{Q}|| := \frac{1}{2} \sum_{\alpha \in \mathcal{A}} |\mathbb{P}(\alpha) - \mathbb{Q}(\alpha)|.$$

Similarly, if $X$ and $Y$ are two random variables ranging over a finite set, their total variation distance, denoted $||X \; ; \; Y||$, is defined as the total variation distance between their distributions.

We also define the *Translated Poisson distribution* as follows.

**Definition 2.1 ([26])** We say that an integer random variable $Y$ has a *translated Poisson distribution* with parameters $\mu$ and $\sigma^2$ and write

$$\mathcal{L}(Y) = TP(\mu, \sigma^2)$$

if $\mathcal{L}(Y - \lfloor \mu - \sigma^2 \rfloor) = Poisson(\sigma^2 + \{\mu - \sigma^2\})$, where $\{\mu - \sigma^2\}$ represents the fractional part of $\mu - \sigma^2$.

Finally, for a positive integer $\ell$, we denote $[\ell] := \{1, \ldots, \ell\}$.

## 3 Statement of Results

### 3.1 A Probabilistic Lemma

We show the following probabilistic lemma, whose proof is given in Sections 4–6.

**Theorem 3.1** Let $\{p_i\}_{i=1}^n$ be arbitrary probability values, $p_i \in [0,1]$ for all $i = 1, \ldots, n$, let $\{X_i\}_{i=1}^n$ be independent indicator random variables such that $X_i$ has expectation $\mathcal{E}[X_i] = p_i$, and let $k$ be a positive integer. Then there exists another set of probabilities $\{q_i\}_{i=1}^n$, $q_i \in [0,1]$, $i = 1, \ldots, n$, which satisfy the following properties:

1. if $\{Y_i\}_{i=1}^n$ are independent indicator random variables such that $Y_i$ has expectation $\mathcal{E}[Y_i] = q_i$, then,

$$\left\| \sum_i X_i \; ; \; \sum_i Y_i \right\| = O(1/k), \qquad (1)$$

$$\text{and, for all } j = 1, \ldots, n, \; \left\| \sum_{i \neq j} X_i \; ; \; \sum_{i \neq j} Y_i \right\| = O(1/k). \qquad (2)$$

2. the set $\{q_i\}_{i=1}^n$ is such that:

   (a) if $p_i = 0$ then $q_i = 0$;

   (b) one of the following is true:

   i. there exists $S \subseteq [n]$ and some value $q$ which is an integer multiple of $\frac{1}{kn}$, such that, for all $i \notin S$, $q_i \in \{0, 1\}$, and, for all $i \in S$, $q_i = q$;

   ii. or, there exists $S \subset [n]$, $|S| < k^3$ such that, for all $i \notin S$, $q_i \in \{0, 1\}$, and, for all $i \in S$, $q_i$ is an integer multiple of $\frac{1}{k^2}$.

## 3.2 Efficient PTAS

From Theorem 3.1, our main result follows. Its proof is based on the following observation: if we replace a strategy profile $(p_i)_{i=1}^n$ that is a Nash equilibrium by the nearby strategy profile $(q_i)_{i=1}^n$ specified by Theorem 3.1, then the change in each player's utility is bounded by the total variation distance between the number of players playing their second strategy in the two strategy profiles. It follows that the search for approximate Nash equilibria can be restricted to the strategy profiles of the form 2(b)i or 2(b)ii specified in Theorem 3.1, and this search can be done efficiently with dynamic programming.

**Theorem 3.2** For any $\epsilon < 1$, an $\epsilon$-Nash equilibrium of a two-strategy anonymous game with $n$ players can be computed in time
$$\text{poly}(n) \cdot U \cdot (1/\epsilon)^{O(1/\epsilon^2)},$$
where $U$ is the number of bits required to represent a payoff value of the game.

*Proof of Theorem 3.2:* Consider a mixed Nash equilibrium $(p_1, \ldots, p_n)$ of the game. We claim that any mixed strategy profile $(q_1, \ldots, q_n)$ satisfying Properties 1 and 2 of Theorem 3.1 is a $O(1/k)$-Nash equilibrium. Indeed, for every player $i \in [n]$ and every pure strategy $s \in \{1, 2\}$ for that player, let us track down the change in the expected utility of the player for playing strategy $s$ when the distribution over $\{0, \ldots, n-1\}$ defined by the probability values $\{p_j\}_{j \neq i}$ is replaced by the distribution defined by the probability values $\{q_j\}_{j \neq i}$. It is not hard to see that the absolute change is bounded by the total variation distance between the distributions of the random variables $\sum_{j \neq i} X_j$ and $\sum_{j \neq i} Y_j$, where $\{X_j\}_{j \neq i}$ are independent Bernoulli random variables with expectations $\mathcal{E}[X_j] = p_j$, $j \neq i$, and $\{Y_j\}_{j \neq i}$ are independent Bernoulli random variables with expectations $\mathcal{E}[Y_j] = q_j$, $j \neq i$;[3] and this total variation distance is $O(1/k)$ by Theorem 3.1. This holds for every choice of $i$ and $s$ and, since the $q_i$'s also satisfy Property 2a of Theorem 3.1, we have that the $q_i$'s constitute an $O(1/k)$-Nash equilibrium of the game. If we take $k = O(1/\epsilon)$, this is an $\epsilon$-Nash equilibrium.

From the previous discussion it follows that there exists a mixed strategy profile $\{q_i\}_i$ which is of the very special kind described by Property 2b in the statement of Theorem 3.1 and constitutes an $\epsilon$-Nash equilibrium of the given game, if we choose $k = O(1/\epsilon)$. The problem is, of course, that we do not know such a mixed strategy profile $\{q_i\}_i$ and we also do not know whether it is of the kind specified by Property 2(b)i or of the kind specified by Property 2(b)ii. Moreover, we cannot afford to do exhaustive search over all mixed strategy profiles satisfying Property 2(b)i or 2(b)ii, since there is an exponential number of those. We do instead the following two searches, corresponding to each of the possibilities 2(b)i and 2(b)ii; one of these searches is guaranteed to find an $\epsilon$-Nash equilibrium.

*Search corresponding to 2(b)i:* We can guess first the cardinality $m$ of the set $S$ (at most $n$ choices), the value $q$ ($kn + 1$ choices), and the number $m'$ of $q_i$'s in $[n] \setminus S$ which are equal to 1 (at most $n$ choices). Then we only need to determine if there exists a set of players $S \subseteq [n]$ and another set of players $S' \subseteq [n] \setminus S$ such that, if all players in $S$ are assigned mixed strategy $q$, all players in $S'$ mixed strategy 1 and all players in $[n] \setminus S \setminus S'$ mixed strategy 0, the resulting mixed strategy profile is an $\epsilon$-Nash equilibrium. To answer this question it is enough to solve the following *max-flow* problem. Let us define the constants $\theta_0 = n - m - m'$, $\theta_q = m$ and $\theta_1 = m'$, and let us consider the bipartite graph $([n], \{0, q, 1\}, E)$ with edge set $E$ defined as follows: $(i, \sigma) \in E$, for $i \in [n]$ and $\sigma \in \{0, q, 1\}$, if $\theta_\sigma > 0$ and $\sigma$ is an $\epsilon$-best response for player $i$, if the partition of the other players into the mixed strategies 0, $q$ and 1 is the partition $\theta$, with one unit subtracted from $\theta_\sigma$.[4] Note that to define $E$ expected payoff computations are required. By straightforward dynamic programming, the expected utility of a player $i$ for playing pure strategy $s \in \{1, 2\}$ given the mixed strategies of the other players can be computed with $O(n^2)$ operations on numbers with at most $b(n, k) := \lceil 1 + n \log_2(kn) + U \rceil$ bits, where $U$ is the number of bits required to specify a payoff value of the game.[5] To conclude the construction of the max-flow instance we add a source node $u$ connected to all the left hand side nodes and a sink node $v$ connected to all the right hand side nodes. We set the capacity of the edge $(\sigma, v)$ equal to $\theta_\sigma$, for all $\sigma \in \{0, q, 1\}$, and the capacity of all other edges equal to 1. If the max-flow from $u$ to $v$ has value $n$ then there is a way to select $S$ and $S'$ by looking at the edges used by the flow (details omitted). There are $O(n^3 k)$ possible guesses for $(m, q, m')$; hence, the search takes overall time

$$O\left((n^3 b(n,k) + p_1(n)) \cdot n^3 k\right), \tag{3}$$

where $p_1(n)$ is the time needed to solve a max-flow problem on a graph with $n + 5$ nodes, $O(n)$ edges, and edge-weights with at most $\lceil \log_2 n \rceil$ bits.

*Search corresponding to 2(b)ii:* We can guess the cardinality $m$ of the set $S$ (there are $k^3$ choices), the number $m'$ of $q_i$'s in $[n] \setminus S$ which are equal to 1 (at most $n$ choices), and a partition of $|S|$ into the integer multiples of $\frac{1}{k^2}$ in $[0, 1]$; let $\{\phi_{i/k^2}\}_{i \in [k^2] \cup \{0\}}$ be the partition. Then we only need to determine if there exists a set of players $S \subseteq [n]$ of cardinality $m$, a set of players $S' \subseteq [n] \setminus S$ of cardinality $m'$, and an assignment of mixed strategies to the players in $S$ with $\phi_{i/k^2}$ of them playing mixed strategy $i/k^2$, so that, if additionally the players in $S'$ are assigned mixed strategy 1 and the players in $[n] \setminus S \setminus S'$ are assigned mixed strategy 0, then the corresponding mixed strategy profile is an $\epsilon$-Nash equilibrium. We answer this question in the same way we did for the case 2(b)i, i.e., by mapping the problem to a max-flow instance. Let us define the vector $\{\theta_{i/k^2}\}_{i \in [k^2] \cup \{0\}}$ by setting $\theta_{i/k^2} = \phi_{i/k^2}$ for all $i \neq 0, 1$, $\theta_{0/k^2} = \phi_{0/k^2} + n - m - m'$ and $\theta_{k^2/k^2} = \phi_{k^2/k^2} + m'$. The bipartite graph of the max-flow instance is now $([n], \{i/k^2\}_{i \in [k^2] \cup \{0\}}, E)$ with edge set $E$ defined as follows: $(j, \sigma) \in E$, for $j \in [n]$ and $\sigma \in \{i/k^2\}_{i \in [k^2] \cup \{0\}}$, if $\theta_\sigma > 0$ and $\sigma$ is an $\epsilon$-best response for player $j$, if the partition of the other players into the mixed strategies $i/k^2$ is the partition $\theta$, with one unit subtracted from $\theta_\sigma$; as argued before, this computation can be carried out with $O(n^2)$ operations on numbers with at most $b'(n, k) := \lceil 1 + n \log(k^2) + U \rceil$ bits.[6] To conclude the construction we include a source node $u$ connected to all left hand side nodes with edges of capacity 1 and a sink node $v$ connected to all the right hand side nodes with edges of capacity $\theta_{i/k^2}$, $i \in [k^2] \cup \{0\}$. If the max-flow from $u$ to $v$ has value $n$ then there is a way to select $S$, $S'$ and $\phi$ by looking at the edges used by the flow (details omitted). There are at most $k^3 \cdot n \cdot ((k+1)e)^{k^2}$ choices for $(m, m', \phi)$; hence, the search takes overall time

$$O\left( \left( n(k^2+1)n^2 b'(n,k) + p_2(n,k) \right) \cdot nk^3((k+1)e)^{k^2} \right), \quad (4)$$

where $p_2(n, k)$ is the time needed to solve a max-flow problem on a graph with $n + k^2 + 3$ nodes, at most $(n+1)(k^2+2) - 1$ edges, and edge-weights with at most $\lceil \log_2 n \rceil$ bits.

---

[3] Recall that all utilities have been normalized to take values in $[0, 1]$.

[4] For our discussion, a mixed strategy $\sigma$ of a player $i$ is an $\epsilon$-*best response* to a set of mixed strategies for the other players iff the expected payoff of player $i$ for playing any pure strategy $s$ in the support of $\sigma$ is no more than $\epsilon$ worse than her expected payoff for playing any pure strategy $s'$.

[5] To compute a bound on the number of bits required for the expected utility computations, note that every non-zero probability value that is computed along the execution of the algorithm must be an integer multiple of $(\frac{1}{kn})^{n-1}$, since the mixed strategies of all players are from the set $\{0, q, 1\}$. Further note that the expected utility is a weighted sum of $(n - 1)$ payoff values, with $U$ bits required to represent each value, and all weights being probabilities.

From (3) and (4), it follows that the overall required time is at most

$$\text{poly}(n) \cdot U \cdot (1/\epsilon)^{O(1/\epsilon^2)}.$$

∎

## 4 Overview of the Proof of Theorem 3.1

We employ a *hybrid argument*. In particular, we define first a set of probability values $\{p'_i\}_{i \in [n]}$ and a corresponding set of independent Bernoulli random variables $\{Z_i\}_{i \in [n]}$ with expectations $\mathcal{E}[Z_i] = p'_i$, for all $i \in [n]$, such that

$$\left\| \sum_i X_i \; ; \; \sum_i Z_i \right\| = O(1/k), \quad (5)$$

and, moreover, for all $j = 1, \ldots, n$, $\left\| \sum_{i \neq j} X_i \; ; \; \sum_{i \neq j} Z_i \right\| = O(1/k). \quad (6)$

The set of probability values $\{p'_i\}_{i \in [n]}$ does not necessarily satisfy Property 2(b)i or 2(b)ii in the statement of Theorem 3.1, but will allow us to define the set of probabilities $\{q_i\}_{i \in [n]}$ which *does*

---

[6] The bound on the number of bits follows from the fact that every non-zero probability value that is computed along the execution of the algorithm must be an integer multiple of $(\frac{1}{k^2})^{n-1}$, since the mixed strategies of all players are from the set $\{i/k^2\}_{i \in [k^2] \cup \{0\}}$. Further, note that the expected utility is a weighted sum of $(n-1)$ payoff values, with $U$ bits required to represent each value, and all weights being probabilities which are multiples of $(\frac{1}{k^2})^{n-1}$.

satisfy Property 2(b)i or 2(b)ii and, moreover,

$$\left\| \sum_i Z_i \; ; \; \sum_i Y_i \right\| = O(1/k), \tag{7}$$

and, for all $j = 1, \ldots, n$, $\left\| \sum_{i \neq j} Z_i \; ; \; \sum_{i \neq j} Y_i \right\| = O(1/k). \tag{8}$

By the triangle inequality, (5) and (7) imply (1), and (6) and (8) imply (2). Let us call *Stage 1* the process of determining the $p'_i$'s and *Stage 2* the process of determining the $q_i$'s. The two stages are described briefly below, and in detail in Sections 5 and 6 respectively.

**Stage 1:** The goal of this stage is to eliminate any probability value $p_i$ falling in the set $\mathcal{T}_k := (0, \frac{1}{k}) \cup (1 - \frac{1}{k}, 1)$, that is, any $p_i$ that is either too small, but non-zero, or too large, but not one. To remove the small $p_i$'s we round some of them to 0 and some of them to $1/k$ in such a way that their sum changes by at most $1/k$. Similarly, we round some of the large $p_i$'s to $1 - 1/k$ and some of them to 1, so that their sum changes by at most $1/k$. Finally, we leave the $p_i$'s falling outside $\mathcal{T}_k$ unchanged. Our work from [14] implies that the set of probability values $\{p'_i\}_i$ thus defined satisfies (5) and (6). See details in Section 5.

**Stage 2:** The definition of the set $\{q_i\}_i$ depends on the number $m$ of $p'_i$'s which are different than 0 and 1. The case $m \geq k^3$ corresponds to the Case 2(b)i in the statement of Theorem 3.1, and the case $m < k^3$ corresponds to the Case 2(b)ii. In both cases we set $q_i = p'_i$, if $p'_i \in \{0, 1\}$; and here is how we round the $p'_i$'s from the index set $\mathcal{M} := \{i \,|\, p'_i \notin \{0, 1\}\}$:

- Case $m \geq k^3$: Using results from the literature on Stein's method, we show that the sum of Bernoulli random variables with expectations $p'_i$, $i \in \mathcal{M}$, can be approximated by a Binomial distribution $B(m', q)$, where $m' \leq m$ and $q$ is an integer multiple of $\frac{1}{kn}$. In particular, we show that an appropriate choice of $m'$ and $q$ implies (7) and (8), if we set $m'$ of the $q_i$'s from the index set $\mathcal{M}$ equal to $q$ and the remaining equal to 0.

- Case $m < k^3$: The Binomial approximation may not be tight enough for small values of $m$. To remedy this, we follow our rounding scheme from [14]. That is, we delicately round the $p'_i$'s to nearby multiples of $\frac{1}{k^2}$ so that their sum $\sum_{i \in \mathcal{M}} p'_i$ is approximated to within $1/k^2$ by the sum $\sum_{i \in \mathcal{M}} q_i$. Our results from [14] imply that (7) and (8) hold in this case.

## 5 Details of Stage 1

We describe the definition of the set $\{p'_i\}_i$. For concreteness, let

$$\mathcal{L} := \{i \,|\, i \in [n] \wedge p_i \in (0, 1/k)\} \quad \text{and} \quad \mathcal{H} := \{i \,|\, i \in [n] \wedge p_i \in (1 - 1/k, 1)\}.$$

We set $p'_i = p_i$, for all $i \in [n] \setminus \mathcal{L} \cup \mathcal{H}$; that is, we leave the probabilities $p_i$ falling outside the set $\mathcal{T}_k$ unchanged. It follows that

$$\left\| \sum_{i \in [n] \setminus \mathcal{L} \cup \mathcal{H}} X_i \; ; \; \sum_{i \in [n] \setminus \mathcal{L} \cup \mathcal{H}} Z_i \right\| = 0, \tag{9}$$

and, for all $j \in [n] \setminus \mathcal{L} \cup \mathcal{H}$, $\left\| \sum_{i \in [n] \setminus (\mathcal{L} \cup \mathcal{H}) \setminus \{j\}} X_i \; ; \; \sum_{i \in [n] \setminus (\mathcal{L} \cup \mathcal{H}) \setminus \{j\}} Z_i \right\| = 0. \tag{10}$

To round the probabilities $p_i$, $i \in \mathcal{L}$, we use the following procedure:

1. Let $S_\mathcal{L} := \sum_{i \in \mathcal{L}} p_i$; $m = \left\lfloor \frac{S_\mathcal{L}}{1/k} \right\rfloor$.

2. Let $\mathcal{L}' \subseteq \mathcal{L}$ be an arbitrary subset of $\mathcal{L}$ with cardinality $|\mathcal{L}'| = m$.

3. Set $p'_i = \frac{1}{k}$, for all $i \in \mathcal{L}'$, and $p'_i = 0$, for all $i \in \mathcal{L} \setminus \mathcal{L}'$.

An application of Lemma 3.9 from [14] with $\alpha = 1$ implies immediately that

$$\left\| \sum_{i \in \mathcal{L}} X_i \; ; \; \sum_{i \in \mathcal{L}} Z_i \right\| \leq \frac{3}{k}, \tag{11}$$

and, for all $j \in \mathcal{L}$, $\left\| \sum_{i \in \mathcal{L} \setminus \{j\}} X_i \; ; \; \sum_{i \in \mathcal{L} \setminus \{j\}} Z_i \right\| \leq \frac{6}{k}, \tag{12}$

where we used that $\left| \sum_{i \in \mathcal{L}} p_i - \sum_{i \in \mathcal{L}} p'_i \right| \leq \frac{1}{k}$ and, for all $j \in \mathcal{L}$, $\left| \sum_{i \in \mathcal{L} \setminus \{j\}} p_i - \sum_{i \in \mathcal{L} \setminus \{j\}} p'_i \right| \leq \frac{2}{k}$.

We follow a similar rounding scheme for the probabilities $p_i$, $i \in \mathcal{H}$; that is, we round some to $1 - 1/k$ and some to $1$ in such a way that their sum is preserved to within $1/k$. As a result, we get (to see this, repeat the argument we employed above to the variables $1 - X_i$ and $1 - Z_i$, $i \in \mathcal{H}$)

$$\left\| \sum_{i \in \mathcal{H}} X_i \; ; \; \sum_{i \in \mathcal{H}} Z_i \right\| \leq \frac{3}{k}, \tag{13}$$

and, for all $j \in \mathcal{H}$, $\left\| \sum_{i \in \mathcal{H} \setminus \{j\}} X_i \; ; \; \sum_{i \in \mathcal{H} \setminus \{j\}} Z_i \right\| \leq \frac{6}{k}. \tag{14}$

Using (9), (10), (11), (12), (13), (14) and the coupling lemma we get (5) and (6).

## 6 Details of Stage 2

Recall that $\mathcal{M} := \{i \mid p'_i \notin \{0, 1\}\}$ and $m := |\mathcal{M}|$. The definition of the probability values $\{q_i\}_i$ will depend on whether $m \geq k^3$ or $m < k^3$. In particular, the case $m \geq k^3$ will correspond to the Case 2(b)i in the statement of Theorem 3.1, and the case $m < k^3$ will correspond to the Case 2(b)ii. In both cases we set

$$q_i = p'_i, \quad \text{for all } i \in [n] \setminus \mathcal{M}.$$

It follows that

$$\left\| \sum_{i \in [n] \setminus \mathcal{M}} Z_i \; ; \; \sum_{i \in [n] \setminus \mathcal{M}} Y_i \right\| = 0, \tag{15}$$

and, for all $j \in [n] \setminus \mathcal{M}$, $\left\| \sum_{i \in [n] \setminus \mathcal{M} \setminus \{j\}} Z_i \; ; \; \sum_{i \in [n] \setminus \mathcal{M} \setminus \{j\}} Y_i \right\| = 0. \tag{16}$

## 6.1 The Case $m \geq k^3$

We show that the random variable $\sum_{i \in \mathcal{M}} Z_i$ is within total variation distance $O(1/k)$ from a Binomial distribution $B(m', q)$ with

$$m' := \left\lceil \frac{(\sum_{i \in \mathcal{M}} p'_i)^2}{\sum_{i \in \mathcal{M}} p'^2_i} \right\rceil \quad \text{and} \quad q := \frac{\ell^*}{kn},$$

where $\ell^* \in \{0, \ldots, kn\}$ satisfies $\frac{\sum_{i \in \mathcal{M}} p'_i}{m'} \in [\frac{\ell^*}{kn}, \frac{\ell^*+1}{kn}]$.

In particular, let us choose an arbitrary subset $\mathcal{M}' \subseteq \mathcal{M}$ with cardinality $m'$ (note that $m' \leq m$ by the Cauchy-Schwarz inequality) and let us set

$$q_i = q, \text{ for all } i \in \mathcal{M}',$$
$$q_i = 0, \text{ for all } i \in \mathcal{M} \setminus \mathcal{M}'.$$

We shall now compare the distributions of the random variables $\sum_{i \in \mathcal{M}} Z_i$ and $\sum_{i \in \mathcal{M}} Y_i$. For this let us set

$$\mu := \mathcal{E}\left[\sum_{i \in \mathcal{M}} Z_i\right] \quad \text{and} \quad \mu' := \mathcal{E}\left[\sum_{i \in \mathcal{M}} Y_i\right],$$

$$\sigma^2 := \text{Var}\left[\sum_{i \in \mathcal{M}} Z_i\right] \quad \text{and} \quad \sigma'^2 := \text{Var}\left[\sum_{i \in \mathcal{M}} Y_i\right].$$

The following lemma compares the values $\mu$, $\mu'$, $\sigma$, $\sigma'$.

**Lemma 6.1** *The following hold*

$$|\mu - \mu'| \leq \frac{1}{k}, \tag{17}$$

$$|\sigma'^2 - \sigma^2| \leq 1 + \frac{3}{k}, \tag{18}$$

$$\mu \geq k^2, \tag{19}$$

$$\sigma^2 \geq k^2\left(1 - \frac{1}{k}\right). \tag{20}$$

The proof of Lemma 6.1 is given in the appendix. To compare $\sum_{i \in \mathcal{M}} Z_i$ and $\sum_{i \in \mathcal{M}} Y_i$ we approximate both by Translated Poisson distributions. To do this, we make use of the following theorem, due to Röllin [26].

**Theorem 6.2 ([26])** *Let $J_1, \ldots, J_n$ be a sequence of independent random indicators with $\mathcal{E}[J_i] = p_i$. Then*

$$\left\|\sum_{i=1}^n J_i \; ; \; TP(\mu, \sigma^2)\right\| \leq \frac{\sqrt{\sum_{i=1}^n p_i^3(1-p_i)} + 2}{\sum_{i=1}^n p_i(1-p_i)},$$

*where $\mu = \sum_{i=1}^n p_i$ and $\sigma^2 = \sum_{i=1}^n p_i(1-p_i)$.*

Theorem 6.2 implies that

$$\left\| \sum_{i \in \mathcal{M}} Z_i \; ; \; TP(\mu, \sigma^2) \right\| \leq \frac{\sqrt{\sum_{i \in \mathcal{M}} p_i'^3 (1 - p_i')} + 2}{\sum_{i \in \mathcal{M}} p_i'(1 - p_i')} \leq \frac{\sqrt{\sum_{i \in \mathcal{M}} p_i'(1 - p_i')} + 2}{\sum_{i \in \mathcal{M}} p_i'(1 - p_i')}$$

$$\leq \frac{1}{\sqrt{\sum_{i \in \mathcal{M}} p_i'(1 - p_i')}} + \frac{2}{\sum_{i \in \mathcal{M}} p_i'(1 - p_i')} = \frac{1}{\sigma} + \frac{2}{\sigma^2}$$

$$\leq \frac{1}{k\sqrt{1 - 1/k}} + \frac{2}{k^2 \left(1 - \frac{1}{k}\right)} = O(1/k). \qquad \text{(using (20))}$$

Similarly,

$$\left\| \sum_{i \in \mathcal{M}} Y_i \; ; \; TP(\mu', \sigma'^2) \right\| \leq \frac{1}{\sigma'} + \frac{2}{\sigma'^2}$$

$$\leq \frac{1}{k\sqrt{1 - \frac{1}{k} - \frac{1}{k^2} - \frac{3}{k^3}}} + \frac{2}{k^2 \left(1 - \frac{1}{k} - \frac{1}{k^2} - \frac{3}{k^3}\right)} = O(1/k). \quad \text{(using (18),(20))}$$

By the triangle inequality we then have that

$$\left\| \sum_{i \in \mathcal{M}} Z_i \; ; \; \sum_{i \in \mathcal{M}} Y_i \right\| \leq \left\| \sum_{i \in \mathcal{M}} Z_i \; ; \; TP(\mu, \sigma^2) \right\| + \left\| \sum_{i \in \mathcal{M}} Y_i \; ; \; TP(\mu', \sigma'^2) \right\| + \left\| TP(\mu, \sigma^2) \; ; \; TP(\mu', \sigma'^2) \right\|$$

$$= O(1/k) + \left\| TP(\mu, \sigma^2) \; ; \; TP(\mu', \sigma'^2) \right\|. \tag{21}$$

It remains to bound the total variation distance between the two Translated Poisson distributions. We make use of the following lemma.

**Lemma 6.3 ([2])** *Let $\mu_1, \mu_2 \in \mathbb{R}$ and $\sigma_1^2, \sigma_2^2 \in \mathbb{R}_+ \setminus \{0\}$ be such that $\lfloor \mu_1 - \sigma_1^2 \rfloor \leq \lfloor \mu_2 - \sigma_2^2 \rfloor$. Then*

$$\left\| TP(\mu_1, \sigma_1^2) - TP(\mu_2, \sigma_2^2) \right\| \leq \frac{|\mu_1 - \mu_2|}{\sigma_1} + \frac{|\sigma_1^2 - \sigma_2^2| + 1}{\sigma_1^2}.$$

Lemma 6.3 implies

$$\left\| TP(\mu, \sigma^2) \; ; \; TP(\mu', \sigma'^2) \right\| \leq \frac{|\mu - \mu'|}{\min(\sigma, \sigma')} + \frac{|\sigma^2 - \sigma'^2| + 1}{\min(\sigma^2, \sigma'^2)}$$

$$\leq \frac{1/k}{k\sqrt{1 - \frac{1}{k} - \frac{1}{k^2} - \frac{3}{k^3}}} + \frac{2 + 3/k}{k^2 \left(1 - \frac{1}{k} - \frac{1}{k^2} - \frac{3}{k^3}\right)} \quad \text{(using Lemma 6.1)}$$

$$= O(1/k^2). \tag{22}$$

Using (21) and (22) we get

$$\left\| \sum_{i \in \mathcal{M}} Z_i \; ; \; \sum_{i \in \mathcal{M}} Y_i \right\| = O(1/k). \tag{23}$$

We claim that the following is also true (see proof in the appendix).

**Lemma 6.4** *For all $j \in \mathcal{M}$:*

$$\left\| \sum_{i \in \mathcal{M} \setminus \{j\}} Z_i \; ; \; \sum_{i \in \mathcal{M} \setminus \{j\}} Y_i \right\| = O(1/k). \tag{24}$$

## 6.2 The Case $m < k^3$

Theorem 3.1 of [14] implies that there exists a set of probability values $\{q_i\}_{i \in \mathcal{M}}$, such that

- $q_i$ is an integer multiple of $\frac{1}{k^2}$, for all $i \in \mathcal{M}$;

$$\left\| \sum_{i \in \mathcal{M}} Z_i \; ; \; \sum_{i \in \mathcal{M}} Y_i \right\| = O(1/k); \tag{25}$$

- and, for all $j \in \mathcal{M}$, $\left\| \sum_{i \in \mathcal{M} \setminus \{j\}} Z_i \; ; \; \sum_{i \in \mathcal{M} \setminus \{j\}} Y_i \right\| = O(1/k). \tag{26}$

## 6.3 Concluding Stage 2

In the case $m \geq k^3$ considered in Section 6.1, a set of probability values $\{q_i\}_i$ was defined which satisfied Property 2(b)i in the statement of Theorem 3.1. In the case $m < k^3$ considered in Section 6.2, the resulting set $\{q_i\}_i$ satisfied Property 2(b)ii. Moreover, in both cases, the following were satisfied

$$\left\| \sum_{i \in \mathcal{M}} Z_i \; ; \; \sum_{i \in \mathcal{M}} Y_i \right\| = O(1/k) \tag{27}$$

$$\left\| \sum_{i \in [n] \setminus \mathcal{M}} Z_i \; ; \; \sum_{i \in [n] \setminus \mathcal{M}} Y_i \right\| = 0, \tag{28}$$

and,

$$\text{for all } j \in \mathcal{M}, \left\| \sum_{i \in \mathcal{M} \setminus \{j\}} Z_i \; ; \; \sum_{i \in \mathcal{M} \setminus \{j\}} Y_i \right\| = O(1/k), \tag{29}$$

$$\text{for all } j \in [n] \setminus \mathcal{M}, \left\| \sum_{i \in [n] \setminus \mathcal{M} \setminus \{j\}} Z_i \; ; \; \sum_{i \in [n] \setminus \mathcal{M} \setminus \{j\}} Y_i \right\| = 0. \tag{30}$$

Using (27), (28), (29), (30) and the coupling lemma, we get (7) and (8).

**Acknowledgment:** We thank Christos Papadimitriou for helpful discussions.

# APPENDIX

## A  Missing Proofs

*Proof of lemma 6.1:* We have
$$\frac{\sum_{i \in \mathcal{M}} p'_i}{m'} - \frac{1}{kn} \le q = \frac{\ell^*}{kn} \le \frac{\sum_{i \in \mathcal{M}} p'_i}{m'}.$$

Hence,
$$\mu - \mu' = \sum_{i \in \mathcal{M}} p'_i - m'q \le \sum_{i \in \mathcal{M}} p'_i - \sum_{i \in \mathcal{M}} p'_i + \frac{m'}{kn} \le \frac{1}{k}$$

and
$$\mu - \mu' \ge \sum_{i \in \mathcal{M}} p'_i - m' \frac{\sum_{i \in \mathcal{M}} p'_i}{m'} = 0.$$

Moreover,
$$\mu = \sum_{i \in \mathcal{M}} p'_i \ge m \frac{1}{k} \ge k^2,$$

since $m \ge k^3$. For convenience let us denote $\lambda_2 := \sum_{i \in \mathcal{M}} p'^2_i$. Then
$$\sigma^2 = \sum_{i \in \mathcal{M}} p'_i (1 - p'_i) = \mu - \lambda_2,$$
$$\text{and } \sigma'^2 = \sum_{i \in \mathcal{M}} q_i(1 - q_i) = \mu' - \sum_{i \in \mathcal{M}} q_i^2 = \mu' - m'q^2.$$

Hence,
$$\sigma^2 - \sigma'^2 = (\mu - \mu') + (m'q^2 - \lambda_2).$$

Now note that
$$\frac{\mu^2}{\lambda_2} \le \frac{\left(\sum_{i \in \mathcal{M}} p'_i\right)^2}{\sum_{i \in \mathcal{M}} p'^2_i} \le m' \le \frac{\left(\sum_{i \in \mathcal{M}} p'_i\right)^2}{\sum_{i \in \mathcal{M}} p'^2_i} + 1 = \frac{\mu^2}{\lambda_2} + 1$$
$$\text{and } \frac{\mu^2}{m'^2} + \frac{1}{(kn)^2} - \frac{2\mu}{m'kn} \le q^2 \le \frac{\mu^2}{m'^2}.$$

Therefore,
$$\frac{\mu^2}{\frac{\mu^2}{\lambda_2} + 1} - \frac{2\mu}{kn} \le \frac{\mu^2}{m'} + \frac{m'}{(kn)^2} - \frac{2\mu}{kn} \le m'q^2 \le \frac{\mu^2}{m'} \le \frac{\mu^2}{\frac{\mu^2}{\lambda_2}} = \lambda_2.$$

which implies

$$|m'q^2 - \lambda_2| \leq \left|\frac{\mu^2}{\frac{\mu^2}{\lambda_2}+1} - \lambda_2\right| + \frac{2\mu}{kn} \leq \frac{\lambda_2^2}{\mu^2 + \lambda_2} + \frac{2m}{kn} \leq \frac{\lambda_2^2}{\mu^2} + \frac{2}{k} \leq 1 + \frac{2}{k}.$$

Hence,

$$|\sigma^2 - \sigma'^2| \leq |\mu - \mu'| + |m'q^2 - \lambda_2| \leq \frac{1}{k} + \frac{2}{k} + 1 \leq 1 + \frac{3}{k}.$$

Finally,

$$\sigma^2 = \sum_{i \in \mathcal{M}} p'_i(1-p'_i) \geq m\frac{1}{k}\left(1-\frac{1}{k}\right) \geq k^2\left(1-\frac{1}{k}\right).$$

∎

*Proof of lemma 6.4:* Let us fix some $j \in \mathcal{M}$ and denote

$$\mu_{-j} := \mathcal{E}\left[\sum_{i \in \mathcal{M}\setminus\{j\}} Z_i\right] \quad \text{and} \quad \mu'_{-j} := \mathcal{E}\left[\sum_{i \in \mathcal{M}\setminus\{j\}} Y_i\right],$$

$$\sigma^2_{-j} := \text{Var}\left[\sum_{i \in \mathcal{M}\setminus\{j\}} Z_i\right] \quad \text{and} \quad \sigma'^2_{-j} := \text{Var}\left[\sum_{i \in \mathcal{M}\setminus\{j\}} Y_i\right].$$

Using Lemma 6.1 it is easy to see that $\mu_{-j}, \mu'_{-j}, \sigma_{-j}, \sigma'_{-j}$ satisfy the following.

**Lemma A.1**

$$|\mu_{-j} - \mu'_{-j}| \leq 1 + \frac{1}{k}, \tag{31}$$

$$|\sigma'^2_{-j} - \sigma^2_{-j}| \leq 1.25 + \frac{3}{k}, \tag{32}$$

$$\mu_{-j} \geq k^2 - 1, \tag{33}$$

$$\sigma^2_{-j} \geq k^2\left(1-\frac{1}{k}\right) - \frac{1}{4}. \tag{34}$$

*Proof of Lemma A.1:* Note that $\mu_{-j} = \mu - p'_j$ and $\mu'_{-j} = \mu' - q_j$. Hence, using Lemma 6.1 we have

$$\mu_{-j} \geq k^2 - 1 \text{ and } |\mu_{-j} - \mu'_{-j}| \leq |\mu - \mu'| + |p'_j - q_j| \leq \frac{1}{k} + 1.$$

Similarly, $\sigma^2_{-j} = \sigma^2 - p'_j(1-p'_j)$ and $\sigma'^2_{-j} = \sigma'^2 - q_j(1-q_j)$. Hence, using Lemma 6.1 we have

$$\sigma^2_{-j} \geq k^2\left(1-\frac{1}{k}\right) - \frac{1}{4}$$

and $|\sigma'^2_{-j} - \sigma^2_{-j}| \leq |\sigma'^2 - \sigma^2| + |p'_j(1-p'_j) - q_j(1-q_j)| \leq 1.25 + \frac{3}{k}.$

∎

Theorem 6.2 implies that

$$\left\|\sum_{i \in \mathcal{M}\setminus\{j\}} Z_i \; ; \; TP(\mu_{-j}, \sigma^2_{-j})\right\| \leq \frac{1}{\sigma_{-j}} + \frac{2}{\sigma^2_{-j}} \leq \frac{1}{\sqrt{k^2\left(1-\frac{1}{k}\right) - \frac{1}{4}}} + \frac{2}{k^2\left(1-\frac{1}{k}\right) - \frac{1}{4}} \quad \text{(using (34))}$$

$$= O(1/k). \tag{35}$$

Similarly,

$$\left\| \sum_{i \in \mathcal{M} \setminus \{j\}} Y_i \;;\; TP(\mu'_{-j}, \sigma'^2_{-j}) \right\| \leq \frac{1}{\sigma'_{-j}} + \frac{2}{\sigma'^2_{-j}} \quad (36)$$

$$\leq \frac{1}{\sqrt{k^2\left(1 - \frac{1}{k}\right) - 1.5 - \frac{3}{k}}} + \frac{2}{k^2\left(1 - \frac{1}{k}\right) - 1.5 - \frac{3}{k}} \quad \text{(using (32),(34))}$$

$$= O(1/k). \quad (37)$$

Finally, from Lemma 6.3 and Lemma A.1 we have

$$\left\| TP(\mu_{-j}, \sigma^2_{-j}) \;;\; TP(\mu'_{-j}, \sigma'^2_{-j}) \right\| \leq \frac{|\mu_{-j} - \mu'_{-j}|}{\min(\sigma^2_{-j}, \sigma'^2_{-j})} + \frac{|\sigma^2_{-j} - \sigma'^2_{-j}| + 1}{\min(\sigma^2_{-j}, \sigma'^2_{-j})}$$

$$\leq \frac{1 + \frac{1}{k}}{\sqrt{k^2\left(1 - \frac{1}{k}\right) - 1.5 - \frac{3}{k}}} + \frac{2.25 + \frac{3}{k}}{k^2\left(1 - \frac{1}{k}\right) - 1.5 - \frac{3}{k}}$$

$$= O(1/k). \quad (38)$$

By the triangle inequality and (35), (37), (38) we get that

$$\left\| \sum_{i \in \mathcal{M} \setminus \{j\}} Y_i \;;\; \sum_{i \in \mathcal{M} \setminus \{j\}} Z_i \right\| = O(1/k).$$

■